\PassOptionsToPackage{unicode}{hyperref}
\PassOptionsToPackage{hyphens}{url}
\PassOptionsToPackage{dvipsnames,svgnames,x11names}{xcolor}
\documentclass[
]{article}

\usepackage{amsmath,amssymb}
\usepackage{lmodern}
\usepackage{iftex}
\ifPDFTeX
  \usepackage[T1]{fontenc}
  \usepackage[utf8]{inputenc}
  \usepackage{textcomp} 
\else 
  \usepackage{unicode-math}
  \defaultfontfeatures{Scale=MatchLowercase}
  \defaultfontfeatures[\rmfamily]{Ligatures=TeX,Scale=1}
\fi
\IfFileExists{upquote.sty}{\usepackage{upquote}}{}
\IfFileExists{microtype.sty}{
  \usepackage[]{microtype}
  \UseMicrotypeSet[protrusion]{basicmath} 
}{}
\makeatletter
\@ifundefined{KOMAClassName}{
  \IfFileExists{parskip.sty}{%
    \usepackage{parskip}
  }{
    \setlength{\parindent}{0pt}
    \setlength{\parskip}{6pt plus 2pt minus 1pt}}
}{
  \KOMAoptions{parskip=half}}
\makeatother
\usepackage{xcolor}
\ifx\paragraph\undefined\else
  \let\oldparagraph\paragraph
  \renewcommand{\paragraph}[1]{\oldparagraph{#1}\mbox{}}
\fi
\ifx\subparagraph\undefined\else
  \let\oldsubparagraph\subparagraph
  \renewcommand{\subparagraph}[1]{\oldsubparagraph{#1}\mbox{}}
\fi

\usepackage{longtable,booktabs,array}
\usepackage{calc} 
\usepackage{etoolbox}
\makeatletter
\patchcmd\longtable{\par}{\if@noskipsec\mbox{}\fi\par}{}{}
\makeatother
\IfFileExists{footnotehyper.sty}{\usepackage{footnotehyper}}{\usepackage{footnote}}
\makesavenoteenv{longtable}
\usepackage{graphicx}
\makeatletter
\def\maxwidth{\ifdim\Gin@nat@width>\linewidth\linewidth\else\Gin@nat@width\fi}
\def\maxheight{\ifdim\Gin@nat@height>\textheight\textheight\else\Gin@nat@height\fi}
\makeatother
\setkeys{Gin}{width=\maxwidth,height=\maxheight,keepaspectratio}
\makeatletter
\def\fps@figure{htbp}
\makeatother
\newlength{\cslhangindent}
\newlength{\csllabelwidth}
\newlength{\cslentryspacingunit} 
\newenvironment{CSLReferences}[2] 
 {
  \setlength{\parindent}{0pt}
  \ifodd #1
  \let\oldpar\par
  \def\par{\hangindent=\cslhangindent\oldpar}
  \fi
  \setlength{\parskip}{#2\cslentryspacingunit}
 }%
 {}
\usepackage{calc}

\usepackage{booktabs}
\usepackage{longtable}
\usepackage{array}
\usepackage{multirow}
\usepackage{wrapfig}
\usepackage{float}
\usepackage{colortbl}
\usepackage{pdflscape}
\usepackage{tabu}
\usepackage{threeparttable}
\usepackage{threeparttablex}
\usepackage[normalem]{ulem}
\usepackage{makecell}
\usepackage{xcolor}
\usepackage{arxiv}
\usepackage{orcidlink}
\usepackage{amsmath}
\usepackage[T1]{fontenc}
\makeatletter
\@ifpackageloaded{caption}{}{\usepackage{caption}}
\AtBeginDocument{%
\ifdefined\contentsname
  \renewcommand*\contentsname{Table of contents}
\else
  \newcommand\contentsname{Table of contents}
\fi
\ifdefined\listfigurename
  \renewcommand*\listfigurename{List of Figures}
\else
  \newcommand\listfigurename{List of Figures}
\fi
\ifdefined\listtablename
  \renewcommand*\listtablename{List of Tables}
\else
  \newcommand\listtablename{List of Tables}
\fi
\ifdefined\figurename
  \renewcommand*\figurename{Figure}
\else
  \newcommand\figurename{Figure}
\fi
\ifdefined\tablename
  \renewcommand*\tablename{Table}
\else
  \newcommand\tablename{Table}
\fi
}
\@ifpackageloaded{float}{}{\usepackage{float}}
\floatstyle{ruled}
\@ifundefined{c@chapter}{\newfloat{codelisting}{h}{lop}}{\newfloat{codelisting}{h}{lop}[chapter]}
\floatname{codelisting}{Listing}

\makeatother
\makeatletter
\@ifpackageloaded{caption}{}{\usepackage{caption}}
\@ifpackageloaded{subcaption}{}{\usepackage{subcaption}}
\makeatother
\makeatletter
\@ifpackageloaded{tcolorbox}{}{\usepackage[many]{tcolorbox}}
\makeatother
\makeatletter
\@ifundefined{shadecolor}{\definecolor{shadecolor}{rgb}{.97, .97, .97}}
\makeatother
\ifLuaTeX
  \usepackage{selnolig}  
\fi
\IfFileExists{bookmark.sty}{\usepackage{bookmark}}{\usepackage{hyperref}}
\IfFileExists{xurl.sty}{\usepackage{xurl}}{} 
\hypersetup{
  pdftitle={Assessing the performance of spatial cross-validation approaches for models of spatially structured data},
  pdfauthor={Michael J Mahoney; Lucas K Johnson; Julia Silge; Hannah Frick; Max Kuhn; Colin M Beier},
  pdfkeywords={cross-validation, spatial data, machine learning, random
forests, simulation},
  colorlinks=true,
  linkcolor={blue},
  filecolor={Maroon},
  citecolor={Blue},
  urlcolor={Blue},
  pdfcreator={LaTeX via pandoc}}

\renewcommand{\today}{3/13/23}
\newcommand{\runninghead}{A Preprint }
\renewcommand{\runninghead}{Assessing spatial cross-validation
approaches }
\title{Assessing the performance of spatial cross-validation approaches
for models of spatially structured data}
\author{
\textbf{Michael J Mahoney}~\orcidlink{0000-0003-2402-304X}\\Graduate
Program in Environmental Science\\State University of New York College
of Environmental Science and Forestry\\Syracuse,
NY,\ 13210\\\href{mailto:mjmahone@esf.edu}{mjmahone@esf.edu}\\\\\\
\textbf{Lucas K Johnson}~\orcidlink{0000-0002-7953-0260}\\Graduate
Program in Environmental Science\\State University of New York College
of Environmental Science and Forestry\\Syracuse,
NY,\ 13210\\\href{mailto:ljohns11@esf.edu}{ljohns11@esf.edu}\\\\\\
\textbf{Julia Silge}~\orcidlink{0000-0002-3671-836X}\\Posit,
PBC\\Boston,
MA,\ 02210\\\href{mailto:julia.silge@posit.co}{julia.silge@posit.co}\\\\\\
\textbf{Hannah Frick}~\orcidlink{0000-0002-6049-5258}\\Posit,
PBC\\Boston,
MA,\ 02210\\\href{mailto:hannah@posit.co}{hannah@posit.co}\\\\\\
\textbf{Max Kuhn}~\orcidlink{0000-0003-2402-136X}\\Posit, PBC\\Boston,
MA,\ 02210\\\href{mailto:max@posit.co}{max@posit.co}\\\\\\
\textbf{Colin M Beier}~\orcidlink{0000-0003-2692-7296}\\Department of
Sustainable Resources Management\\State University of New York College
of Environmental Science and Forestry\\Syracuse,
NY,\ 13210\\\href{mailto:cbeier@esf.edu}{cbeier@esf.edu}}
\date{3/13/23}
\begin{document}
\maketitle
\begin{abstract}
Evaluating models fit to data with internal spatial structure requires
specific cross-validation (CV) approaches, because randomly selecting
assessment data may produce assessment sets that are not truly
independent of data used to train the model. Many spatial CV
methodologies have been proposed to address this by forcing models to
extrapolate spatially when predicting the assessment set. However, to
date there exists little guidance on which methods yield the most
accurate estimates of model performance.

We conducted simulations to compare model performance estimates produced
by five common CV methods fit to spatially structured data. We found
spatial CV approaches generally improved upon resubstitution and V-fold
CV estimates, particularly when approaches which combined assessment
sets of spatially conjunct observations with spatial exclusion buffers.
To facilitate use of these techniques, we introduce the
\texttt{spatialsample} package which provides tooling for performing
spatial CV as part of the broader tidymodels modeling framework.
\end{abstract}
{\bfseries \emph Keywords}
\def\sep{\textbullet\ }
cross-validation \sep spatial data \sep machine learning \sep random
forests \sep 
simulation

\ifdefined\Shaded\renewenvironment{Shaded}{\begin{tcolorbox}[breakable, borderline west={3pt}{0pt}{shadecolor}, sharp corners, enhanced, interior hidden, frame hidden, boxrule=0pt]}{\end{tcolorbox}}\fi

\hypertarget{sec-introduction}{%
\section{Introduction}\label{sec-introduction}}

Evaluating predictive models fit using data with internal spatial
dependence structures, as is common for Earth science and environmental
data (Legendre and Fortin 1989), is a difficult task. Low-bias models,
such as the machine learning techniques gaining traction across the
literature, may overfit on the data used to train them. As a result,
model performance metrics may be nearly perfect when models are used to
predict the data used to train the model (the ``resubstitution
performance'' of the model) (Kuhn and Johnson 2013), but deteriorate
when presented with new data.

In order to detect a model's failure to generalize to new data, standard
cross-validation (CV) evaluation approaches assign each observation to
one or more ``assessment'' sets, then average performance metrics
calculated against each assessment set using predictions from models fit
using ``analysis'' sets containing all non-assessment observations. We
refer to the assessment set, which is ``held out'' from the model
fitting process, as \(D_{\operatorname{out}}\), while we refer to the
analysis set as \(D_{\operatorname{in}}\). Splitting data between
\(D_{\operatorname{out}}\) and \(D_{\operatorname{in}}\) is typically
done randomly, which is sufficient to determine model performance on
new, unrelated observations when working with data whose variables are
independent and identically distributed. However, when models are fit
using variables with internal spatial dependence (often referred to as
spatial autocorrelation), random assignment will likely assign
neighboring observations to both \(D_{\operatorname{out}}\) and
\(D_{\operatorname{in}}\). Given that neighboring observations are often
more closely related (Legendre and Fortin 1989), this random assignment
yields similar results as the resubstitution performance, providing
over-optimistic validation results that over state the ability of the
model to generalize to new observations or to regions not
well-represented in the training data (Roberts et al. 2017; Bahn and
McGill 2012).

One potential solution for this problem is to not assign data to
\(D_{\operatorname{out}}\) purely at random, but to instead section the
data into ``blocks'' based upon its dependence structure and assign
entire blocks to \(D_{\operatorname{out}}\) as a unit (Roberts et al.
2017). For data with spatial structure, this means assigning
observations to \(D_{\operatorname{out}}\) and \(D_{\operatorname{in}}\)
based upon their spatial location, in order to increase the average
distance between observations in \(D_{\operatorname{out}}\) and those
used to train the model. The amount of distance required to ensure
accurate estimates of model performance is a matter of some debate, with
suggestions to use variously the variogram ranges for model predictors,
model outcomes, or model residuals (Le Rest et al. 2014; Roberts et al.
2017; Telford and Birks 2009; Valavi et al. 2018; Karasiak et al. 2021).
However, it is broadly agreed that increasing the average distance
between observations in \(D_{\operatorname{out}}\) and those used to
train the model may produce more accurate estimates of model
performance.

This objective -- evaluating the performance of a predictive model -- is
subtly but importantly distinct from evaluating the accuracy of a map of
predictions. Map accuracy assessments assume a representative
probability sample in order to produce unbiased accuracy estimates
(Stehman and Foody 2019), while assessments of models fit using spatial
data typically assume a need to estimate model performance without
representative and independent assessment data. Such situations emerge
frequently across model-based studies (Gruijter and Braak 1990; Brus
2020), such as during hyperparameter tuning (Schratz et al. 2019); when
extrapolating spatially to predict into ``unknown space'' without
representative assessment data (Meyer and Pebesma 2021, 2022); or when
working with data collected via non-representative convenience samples
(any non-probability sample where observations are collected on the
easiest to access members of a population) as commonly occurs in ecology
and environmental science (Martin, Blossey, and Ellis 2012; Yates et al.
2018). In these situations, understanding the predictive accuracy of the
model on independent data is essential. Although recent research has
argued against spatial CV for map accuracy assessments (Wadoux et al.
2021), spatial CV remains an essential tool for evaluating the
performance of predictive models.

For many situations spatial CV has been shown to provide empirically
better estimates of model performance than non-spatial methods (Bahn and
McGill 2012; Schratz et al. 2019; Meyer et al. 2018; Le Rest et al.
2014; Ploton et al. 2020), and as such is popularly used to evaluate
applied modeling projects across domains (Townsend, Papeş, and Eaton
2007; Meyer et al. 2019; Adams et al. 2020). As such, a variety of
methods for spatially assigning data to \(D_{\operatorname{out}}\) have
been proposed, many of which have been shown to improve performance
estimates over randomized CV approaches. However, the lack of direct
comparisons between alternative spatial CV approaches makes it difficult
to know which methods may be most effective for estimating model
performance. Understanding the different properties of various CV
methods is particularly important given that many spatial modeling
projects rely on CV for their primary accuracy assessments (Bastin et
al. 2019; Fick and Hijmans 2017; Hoogen et al. 2019; Hengl et al. 2017).

Here, we evaluated the leading spatial CV approaches found in the
literature, using random forest models fit on spatially structured data
following the simulation approach of Roberts et al. (2017). To
facilitate comparison among methods, we offer a useful taxonomy and
definition of these CV approaches, attempting to unify disparate
terminology used throughout the literature. We then provide
comprehensive overview of the performance of these spatial CV methods
across a wide array of parameterizations, providing the first
comparative performance evaluation for many of these techniques. We
found that spatial CV methods yielded overall better estimates of model
performance than random assignment. Approaches that incorporated both
\(D_{\operatorname{out}}\) of spatially conjunct observations and
buffers yielded the best results. Lastly, to facilitate further
evaluation and use of spatial CV methods, we introduce the
\texttt{spatialsample} R package that implements each of the approaches
evaluated here, and avoids many of the pitfalls encountered by prior
implementations.

\hypertarget{spatialsample-and-the-tidymodels-framework}{%
\section{\texorpdfstring{\texttt{spatialsample} and the tidymodels
Framework}{spatialsample and the tidymodels Framework}}\label{spatialsample-and-the-tidymodels-framework}}

The tidymodels framework is a set of open-source packages for the R
programming language that provide a consistent interface for common
modeling tasks across a variety of model families and objectives
following the same fundamental design principles as the tidyverse (R
Core Team 2022; Kuhn and Silge 2022; Wickham et al. 2019). These
packages help users follow best practices while fitting and evaluating
models, with functions and outputs that integrate well with the other
packages in the tidymodels and tidyverse ecosystems as well as with the
rest of the R modeling ecosystem. Historically, data splitting and CV in
the tidymodels framework has been handled by the \texttt{rsample}
package (Frick et al. 2022), with additional components of the
tidymodels ecosystem providing functionality for hyperparameter tuning
and model evaluation (Kuhn 2022; Kuhn and Frick 2022). However,
\texttt{rsample} primarily focuses on randomized CV approaches, and as
such it has historically been difficult to implement spatial CV
approaches within the tidymodels framework.

A new tidymodels package, \texttt{spatialsample}, addresses this gap by
providing a suite of functions to implement the most popular spatial CV
approaches. The resamples created by \texttt{spatialsample} rely on the
same infrastructure as those created by \texttt{rsample}, and as such
can make use of the same tidymodels packages for hyperparameter tuning
and model evaluation. As an implementation of spatial CV methods,
\texttt{spatialsample} improves on alternate implementations in R by
relying on the \texttt{sf} and \texttt{s2} packages for calculating
distance in geographic coordinate reference systems, improving the
accuracy of distance calculations and CV fold assignments when working
with geographic coordinates (Pebesma 2018; Dunnington, Pebesma, and
Rubak 2021). The \texttt{spatialsample} package is also able to
appropriately handle data with coordinate reference systems that use
linear units other than meters and to process user-provided parameters
using any units understood by the \texttt{units} package (Pebesma,
Mailund, and Hiebert 2016). The \texttt{spatialsample} package also
allows users to perform spatial CV procedures while modeling data with
polygon geometries, such as those provided by the US Census Bureau.
Distances between polygons are calculated between polygon edges, rather
than centroids or other internal points, to ensure that adjacent
polygons are handled appropriately by CV functions. Finally,
\texttt{spatialsample} provides a standard interface to apply inclusion
radii and exclusion buffers to all CV methodologies
(Section~\ref{sec-overview}), providing a high degree of flexibility for
specifying spatial CV patterns.

\hypertarget{sec-overview}{%
\section{Resampling Methods}\label{sec-overview}}

This paper evaluates a number of the most popular spatial CV approaches,
based upon their prevalence across the literature and software
implementations. We focus on CV methods which automatically split data
either randomly or spatially but did not address any assessment methods
that divide data based upon pre-specified, user-defined boundaries or
predictor space. As such, we use ``distance'' and related terms to refer
to spatial distances between observations, unless we specifically refer
to ``distance in predictor space''.

\hypertarget{resubstitution}{%
\subsection{Resubstitution}\label{resubstitution}}

Evaluating a model's performance when predicting the same data used to
fit the model yields what is commonly known as the ``apparent
performance'' or ``resubstitution performance'' of the model (Kuhn and
Johnson 2013). This procedure typically produces an overly-optimistic
estimate of model performance, and as such is a poor method for
estimating how well a model will generalize to new data (Efron 1986;
Gong 1986; Efron and Gong 1983). We include an assessment of
resubstitution error in this study for comparison, but do not recommend
it as an evaluation procedure in practice.

\hypertarget{randomized-v-fold-cv}{%
\subsection{Randomized V-fold CV}\label{randomized-v-fold-cv}}

Perhaps the most common approach to CV is V-fold CV, also known as
k-fold CV. In this method, each observation is randomly assigned to one
of \(v\) folds. Models are then fit to each unique combination of
\(v - 1\) folds and evaluated against the remaining fold, with
performance metrics estimated by averaging across the \(v\) iterations
(Stone 1974). V-fold CV is generally believed to be the least biased of
the dominant randomized CV approaches (Kuhn and Johnson 2019), though
research has suggested it overestimates model performance (Varma and
Simon 2006; Bates, Hastie, and Tibshirani 2021). This optimistic bias is
even more notable when training models using data with internal
dependency structures, such as spatial structure (Roberts et al. 2017).

For this study, V-fold CV was performed using the \texttt{vfold\_cv()}
function in \texttt{rsample}.

\hypertarget{blocked-cv}{%
\subsection{Blocked CV}\label{blocked-cv}}

One of the most straightforward forms of spatial CV is to divide the
study area into a set of polygons using a regular grid, with all
observations inside a given polygon assigned to
\(D_{\operatorname{out}}\) as a group (Valavi et al. 2018; Brenning
2012; Wenger and Olden 2012). This technique is known as ``spatial
blocking'' (Roberts et al. 2017) or ``spatial tiling'' (Brenning 2012).
Frequently, each block is used as an independent
\(D_{\operatorname{out}}\) (leave-one-block-out CV, Wenger and Olden
2012), though many implementations allow users to use fewer
\(D_{\operatorname{out}}\), combining multiple blocks into sets either
at random or via a systematic assignment approach (Valavi et al. 2018).
Spatial blocking attempts to address the limitations of randomized
V-fold CV by introducing distance between \(D_{\operatorname{in}}\) and
\(D_{\operatorname{out}}\), though the distance from observations on the
perimeter of a block to \(D_{\operatorname{in}}\) will be much less than
that from observations near the center of the block (O'Sullivan and
Unwin 2010). This disparity can be addressed through the use of an
exclusion buffer around \(D_{\operatorname{out}}\), wherein points
within a certain distance of \(D_{\operatorname{out}}\) (depending on
the implementation, calculated alternatively as distance from the
polygon defining each block, the convex hull of points in
\(D_{\operatorname{out}}\), or from each point in
\(D_{\operatorname{out}}\) independently) are excluded from both
\(D_{\operatorname{out}}\) and \(D_{\operatorname{in}}\) (Valavi et al.
2018).

A challenge with spatial blocking is that dividing the study area using
a standard grid often results in observations in unrelated areas being
grouped together in a single fold, as regular gridlines likely will not
align with relevant environmental features (for instance, blocks may
span significant altitudinal gradients, or reach across a river to
combine disjunct populations). Although this can be mitigated through
careful parameterization of the grid, it is difficult to create
meaningful \(D_{\operatorname{out}}\) while still enforcing the required
spatial separation between \(D_{\operatorname{in}}\) and
\(D_{\operatorname{out}}\).

For this study, spatial blocking was performed using the
\texttt{spatial\_block\_cv()} function in \texttt{spatialsample}. Each
block was treated as a unique fold (leave-one-block-out CV).

\hypertarget{clustered-cv}{%
\subsection{Clustered CV}\label{clustered-cv}}

Another form of spatial CV involves grouping observations into a
pre-specified number of clusters based on their spatial arrangement, and
then treating each cluster as a \(D_{\operatorname{out}}\) in V-fold CV
(Brenning 2012; Walvoort, Brus, and Gruijter 2010). This approach allows
for a great degree of flexibility, as alternative distance calculations
and clustering algorithms may produce substantially different clusters.
Similarly to spatially-blocked CV this approach may produce unbalanced
\(D_{\operatorname{out}}\) and folds that combine unrelated areas,
though in practice most clustering algorithms typically produce more
sensible fold boundaries than spatial blocking. Clustering is also
similar to spatial blocking in that observations closer to the center of
a cluster will be much further spatially separated from
\(D_{\operatorname{in}}\) than those near the perimeter, although
clustering algorithms typically produce more circular tiles than
blocking methods, and as such generally have fewer points along their
perimeter overall. As with spatial blocking, exclusion buffers can be
used to ensure a minimum distance between \(D_{\operatorname{in}}\) and
\(D_{\operatorname{out}}\).

A notable difference between spatial clustering and spatial blocking is
that, depending upon the algorithm used to assign data to clusters,
cluster boundaries may be non-deterministic. This stochasticity means
that repeated CV may be more meaningful with clustered CV than with
spatial blocking, but also makes it difficult to ensure that cluster
boundaries align with meaningful boundaries.

For this study, spatial clustering was performed using the
\texttt{spatial\_clustering\_cv()} function in \texttt{spatialsample}.
Each cluster was treated as a unique fold (leave-one-cluster-out CV).

\hypertarget{buffered-leave-one-observation-out-cv-blo3-cv}{%
\subsection{Buffered Leave-One-Observation-Out CV (BLO3
CV)}\label{buffered-leave-one-observation-out-cv-blo3-cv}}

An alternative approach to spatial CV involves performing leave-one-out
CV, a form of V-fold cross validation where \(v\) is set to the number
of observations such that each observation forms a separate
\(D_{\operatorname{out}}\), with all points within a buffer distance of
\(D_{\operatorname{out}}\) omitted from \(D_{\operatorname{in}}\)
(Telford and Birks 2009; Pohjankukka et al. 2017). As the other methods
investigated here are all examples of leave-one-group-out CV, with
groups defined by spatial positions, we refer to this procedure as
buffered leave-one-observation-out CV (BLO3 CV). This approach may be
more robust to different parameterizations than spatial clustering or
blocking, as the contents of a given \(D_{\operatorname{out}}\) are not
as dependent upon the precise locations of blocking polygons or cluster
boundaries. Many studies have recommended buffered leave-one-out cross
validation for models fit using spatial data, with the size of the
exclusion buffer variously determined by variogram ranges for model
predictors, model outcomes, or model residuals (Le Rest et al. 2014;
Roberts et al. 2017; Telford and Birks 2009; Valavi et al. 2018;
Karasiak et al. 2021).

For this study, BLO3 CV was performed using the
\texttt{spatial\_buffer\_vfold\_cv()} function in
\texttt{spatialsample}.

\hypertarget{leave-one-disc-out-cv-lodo-cv}{%
\subsection{Leave-One-Disc-Out CV (LODO
CV)}\label{leave-one-disc-out-cv-lodo-cv}}

\begin{figure}

{\centering \includegraphics{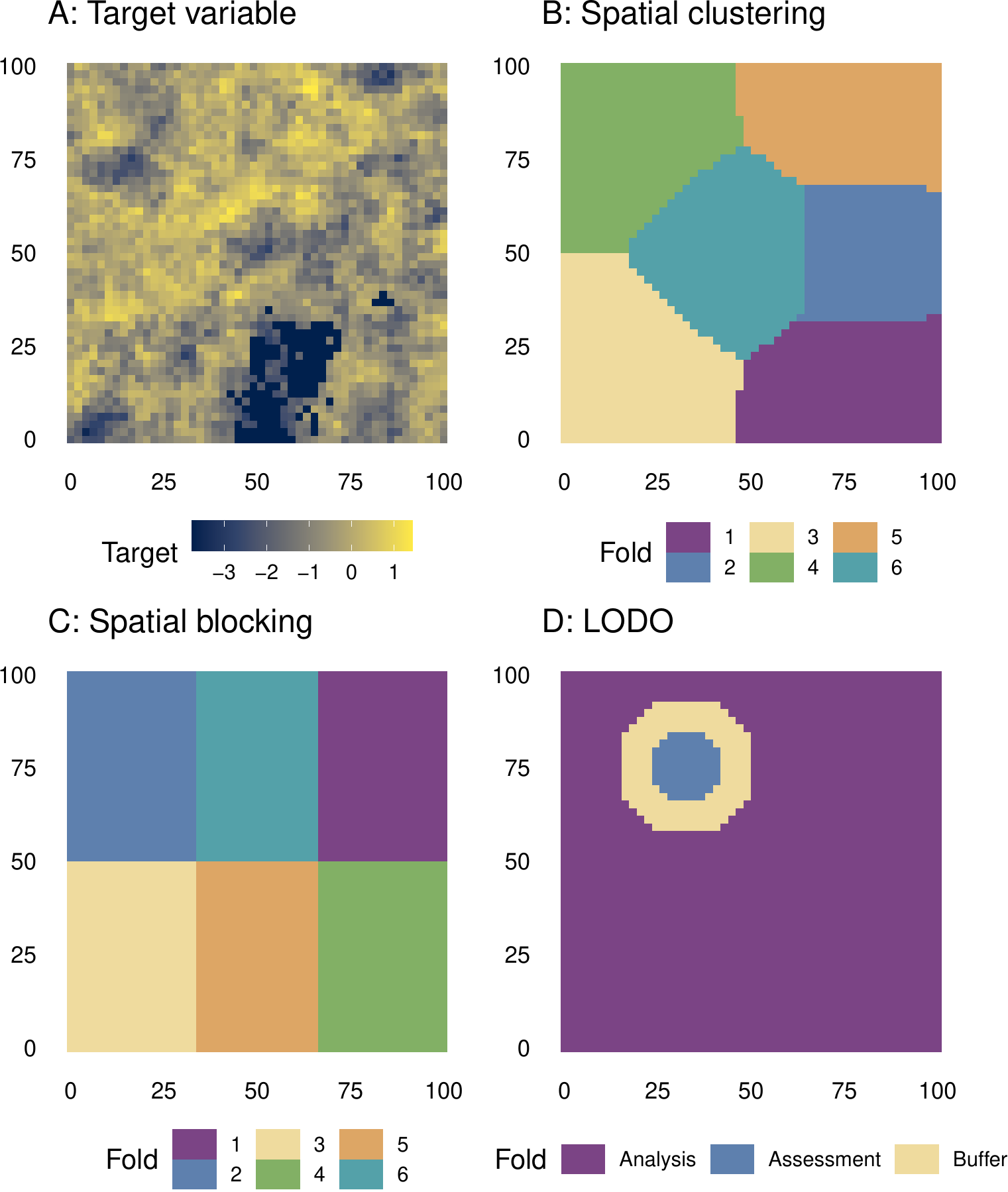}

}

\caption{\label{fig-maps}One simulation of the dependent variable \(y\)
and several of the CV approaches used. A: Simulated data showing
environmental clustering of a variable of interest, named ``Target'', in
one simulated landscape. B: Spatial clustering CV fold assignments, as
produced by \texttt{spatial\_clustering\_cv()}, based upon using k-means
clustering to group observations spatially. C: Spatially blocked CV fold
assignments, produced using \texttt{spatial\_block\_cv()} to group
observations spatially. D: A single fold of LODO CV, showing
\(D_{\operatorname{in}}\) and \(D_{\operatorname{out}}\) as well as the
exclusion buffer, performed using \texttt{spatial\_buffer\_vfold\_cv()}
with a radius and buffer covering 10\% of the mapped area. Points within
the inclusion radius of the randomly selected observation are included
in \(D_{\operatorname{out}}\), while points within the exclusion buffer
of \(D_{\operatorname{out}}\) are not included in either set. For BLO3
and LODO CV, this procedure is repeated for each grid cell.}

\end{figure}

The final spatial CV method investigated here is leave-one-disc-out CV,
following Brenning (2012). This method extends BLO3 by adding all points
within a certain radius of each observation to
\(D_{\operatorname{out}}\), increasing the size of the final
\(D_{\operatorname{out}}\). Data points falling within the exclusion
buffer of any observation in \(D_{\operatorname{out}}\), including those
added by the inclusion radius, is then removed from
\(D_{\operatorname{out}}\) (Figure~\ref{fig-maps}). Similarly to blocked
and clustered CV, LODO CV evaluates models against multiple
\(D_{\operatorname{out}}\) of spatially conjunct observations. Similar
to BLO3 CV, LODO CV approach may be more robust to different parameter
values than methods assigning \(D_{\operatorname{out}}\) based upon
blocking polygons or cluster boundaries. Unlike any of the other
approaches investigated, observations may appear in multiple
\(D_{\operatorname{out}}\), with observations in more intensively
sampled regions being selected more often.

In this study, spatial leave-one-disc-out CV was performed using the
\texttt{spatial\_buffer\_vfold\_cv()} function in
\texttt{spatialsample}. An important feature of this implementation of
leave-disc-out CV is that the exclusion buffer is calculated separately
for each point in \(D_{\operatorname{out}}\). Where other
implementations remove all observations within the buffer distance of
the inclusion radius to create a uniform ``doughnut'' shaped buffer,
\texttt{spatialsample} only removes observations that are within the
buffer distance of data in \(D_{\operatorname{out}}\), potentially
retaining more data in \(D_{\operatorname{in}}\) by creating an
irregular buffer polygon.

\hypertarget{sec-methods}{%
\section{Methods}\label{sec-methods}}

\hypertarget{sec-simulation}{%
\subsection{Landscape Simulation}\label{sec-simulation}}

To compare the validation techniques described above, we extended the
simulation approach used by Roberts et al. (2017, Box 1). We simulated
100 landscapes, representing independent realizations of the same
data-generating process, generating a set of 13 variables calculated
using the same stochastic formulations across a regularly spaced 50 x 50
cell grid, for a total of 2,500 cells per landscape
(Table~\ref{tbl-simulations}). Simulated predictors included eight
random Gaussian fields, generated using the \texttt{RandomFields} R
package (Schlather et al. 2015), which uses stationary isotropic
covariance models to generate spatially structured variables. Five
additional variables were calculated as combinations of the randomly
generated variables to imitate interactions between environmental
variables. This simulation approach was originally designed to resemble
the environmental data that might be used to model species abundance and
distribution; further interpretation of what each predictor represents
is provided in Appendix 2 of Roberts et al. (2017).

\hypertarget{tbl-simulations}{}
\begin{table}
\caption{\label{tbl-simulations}Simulated predictors generated for each independent landscape, created
following Roberts et al. 2017. Predictors are indicated as being used
for y if they were included either in Equation 1 or used to calculate
variables included in Equation 1. Predictors are indicated as being used
in models if they were used as predictors in random forest models. }\tabularnewline

\centering
\begin{tabular}[t]{l>{\raggedright\arraybackslash}p{20em}l>{\raggedright\arraybackslash}p{5em}}
\toprule
Name & Variable Definition & Used for $y$? & Used in model?\\
\midrule
X1 & Random Gaussian field with exponential covariance (variance = 0.1, scale = 0.1) & Yes & No\\
\addlinespace
X2 & Random Gaussian field with exponential covariance (variance = 0.3, scale = 0.1) & No & Yes\\
\addlinespace
X3 & Random Gaussian field with Gaussian covariance (variance = 0.1, scale = 0.3) & No & Yes\\
\addlinespace
X4 & If the ratio (X2 / X3) is above the 95th percentile of all values, 0; else 1. & Yes (excluding) & No\\
\addlinespace
X5 & X1 + X2 + X3 + (X2 $\cdot$ X3) & Yes & No\\
\addlinespace
X6 & Random Gaussian field with exponential covariance (variance = 0.1, scale = 0.1) & Yes & Yes\\
\addlinespace
X7 & Random Gaussian field with exponential covariance (variance = 0.1, scale = 0.1) & No & Yes\\
\addlinespace
X8 & Random Gaussian field with exponential covariance (variance = 0.1, scale = 0.1) & No & Yes\\
\addlinespace
X9 & Random Gaussian field with Gaussian covariance (variance = 0.1, scale = 0.3) & No & Yes\\
\addlinespace
X10 & Random Gaussian field with Gaussian covariance (variance = 0.1, scale = 0.3) & No & Yes\\
\addlinespace
X11 & X2/X3 & Yes (limiting) & No\\
\addlinespace
X12 & $1 / (\operatorname{sqrt}(2\cdot\pi))\cdot\exp-(\operatorname{X3}^2/4)$ & Yes & No\\
\addlinespace
X13 & $1 / (\operatorname{sqrt}(2\cdot\pi))\cdot\exp-(\operatorname{X2}^2/4)$ & Yes & No\\
\bottomrule
\end{tabular}
\end{table}

These predictors were then used to generate a target variable \(y\)
using Equation~\ref{eq-y}.

\begin{equation}\protect\hypertarget{eq-y}{}{
y = 
\left\{
    \begin{array}{lr}
        \min(y), & \text{if } \operatorname{X4} \neq 0\\
        \operatorname{X11}, & \text{if } y\geq \operatorname{X11}\\
        \operatorname{X1} + \operatorname{X5} + \operatorname{X6} + \operatorname{X12} + \operatorname{X13}, & \text{otherwise }
    \end{array}
\right\}
}\label{eq-y}\end{equation}

One instance of the spatially clustered \(y\) values produced by this
process is visualized in Figure~\ref{fig-maps}.

Models were then fit using variables X2, X3, and X6 - X10. Of these
seven variables, three were involved in calculating the target variable
\(y\) (X2 and X3, as components of X4 and X5; and X6, used directly) and
therefore provide useful information for models, while the remaining
four (X7 - X10) were included to allow overfitting.

\hypertarget{sec-resampling}{%
\subsection{Resampling Methodology}\label{sec-resampling}}

We divided each simulated landscape into folds using each of the data
splitting approaches (Section~\ref{sec-overview}) across a wide range of
parameter sets (Table~\ref{tbl-whichparams}; Table~\ref{tbl-paramdefs})
in order to evaluate the usefulness of spatial CV approaches. Spatial
blocking, spatial clustering, and leave-one-disc-out used a
``leave-one-group-out'' approach, where each \(D_{\operatorname{out}}\)
was made up of a single block or cluster of observations, with all other
data (excluding any within the exclusion buffer) used as
\(D_{\operatorname{in}}\). BLO3 used a leave-one-observation-out
approach. We additionally evaluated spatial blocking with fewer
\(D_{\operatorname{out}}\) than blocks, resulting in multiple blocks
being used in each \(D_{\operatorname{out}}\). Each simulated landscape
was resampled independently, meaning that stochastic methods (such as
V-fold CV and spatial clustering) produced different CV folds across
each simulation. All resampling used functions implemented in the
\texttt{rsample} and \texttt{spatialsample} packages (Frick et al. 2022;
Mahoney and Silge 2022). Examples of spatial clustering CV, spatially
blocked CV, and leave-one-disc-out CV are visualized in
Figure~\ref{fig-maps}.

\hypertarget{tbl-whichparams}{}
\begin{table}
\caption{\label{tbl-whichparams}Parameters applied to each CV method assessed, and the number of
iterations performed. 100 iterations were performed per unique
combination of parameters. }\tabularnewline

\centering
\begin{tabular}[t]{l>{}llr}
\toprule
CV Method & Resampling function & Parameters & \# of iterations\\
\midrule
Resubstitution & \ttfamily{} &  & 100\\
\addlinespace
V-fold & \ttfamily{vfold\_cv()} & V & 400\\
\addlinespace
Blocked & \ttfamily{spatial\_block\_cv()} & Block size, Buffer & 8800\\
\addlinespace
Clustered & \ttfamily{spatial\_clustering\_cv()} & V, Buffer, Cluster function & 8800\\
\addlinespace
BLO3 & \ttfamily{spatial\_buffer\_vfold\_cv()} & Buffer & 1700\\
\addlinespace
LODO & \ttfamily{spatial\_buffer\_vfold\_cv()} & Buffer, Radius & 11100\\
\bottomrule
\end{tabular}
\end{table}

\hypertarget{tbl-paramdefs}{}
\begin{table}
\caption{\label{tbl-paramdefs}Definitions of parameters applied to CV methods. }\tabularnewline

\centering
\begin{tabular}[t]{l>{\raggedright\arraybackslash}p{15em}>{\raggedright\arraybackslash}p{20em}}
\toprule
Parameter & Values & Definition\\
\midrule
V & 2, 5, 10, 20; 2, 4, 9, 16, 25, 36, 64, 100 & The number of folds to assign data into. Each fold was used as $D_{\operatorname{out}}$ precisely once. The first set of values were used for spatial clustering, while the second was used for spatial blocking. For spatial clustering, this controls the number of clusters.\\
\addlinespace
Cluster function & K-means, Hierarchical & The algorithm used to cluster observations into folds.\\
\addlinespace
Block size & 1/100, 1/64, 1/36, 1/25, 1/16, 1/9, 1/4, 1/2 & The proportion of the grid each block should occupy (such that 1/2 creates two blocks, each occupying half the grid).\\
\addlinespace
Blocking method & Random, Systematic (continuous), Systematic (snake) & For spatial blocking, the method for assigning blocks to folds: randomly ('random'), in a 'scanline' moving left to right across each row of the grid ('systematic (continuous)'), or moving back and forth across the rows of the grid ('systematic (snake)').\\
\addlinespace
Buffer & 0.00, 0.03, 0.06, 0.09, 0.12, 0.15, 0.18, 0.21, 0.24, 0.27, 0.30, 0.33, 0.36, 0.39, 0.42, 0.45, 0.48 & The size of the exclusion buffer to apply around $D_{\operatorname{out}}$, expressed as a proportion of the side length of the grid. Observations within this distance of any point in $D_{\operatorname{out}}$ are included in neither $D_{\operatorname{in}}$ nor $D_{\operatorname{out}}$. Buffer distances above 0.3 were only used for BLO3 CV, as increased buffer distances around larger $D_{\operatorname{out}}$ may produce empty $D_{\operatorname{in}}$.\\
\addlinespace
Radius & 0.00, 0.03, 0.06, 0.09, 0.12, 0.15, 0.18, 0.21, 0.24, 0.27, 0.30 & The size of the inclusion radius to apply around $D_{\operatorname{out}}$, expressed as a proportion of the side length of the grid. Observations within this distance of any point in $D_{\operatorname{out}}$ are moved from $D_{\operatorname{in}}$ into $D_{\operatorname{out}}$.\\
\bottomrule
\end{tabular}
\end{table}

\hypertarget{sec-models}{%
\subsection{Model Fitting and Evaluation}\label{sec-models}}

For each iteration, we modeled the target variable \(y\) using random
forests as implemented in the \texttt{ranger} R package (Breiman 2001;
Wright and Ziegler 2017), fit using variables X2, X3, and X6 - X10.
Random forests generally provide high predictive accuracy even without
hyperparameter tuning (Probst, Bischl, and Boulesteix 2018), and as such
all random forests were fit using the default hyperparameter settings of
the \texttt{ranger} package, namely 500 decision trees, a minimum of 5
observations per leaf node, and two variables to split on per node.

Model accuracy was measured using root-mean-squared error (RMSE,
Equation~\ref{eq-rmse}). To find the ``ideal'' error rate that we would
expect CV approaches to estimate, we fit 100 separate random forest
models, each trained using all values within one of the 100 simulated
landscapes. We then calculated the RMSE for each of these models when
used to predict each of the 99 other landscapes. As each landscape is an
independent realization of the same data-generation process, the
relationships between predictors and \(y\) is identical across
landscapes, although the spatial relationships between \(y\) and
variables not used to generate \(y\) are likely different across
iterations. As such, RMSE values from a model trained on one landscape
and used to predict the others represent the ability of the model to
predict \(y\) based upon the predictors and without relying upon spatial
structure. These RMSE estimates therefore represent the ``true'' range
of RMSE values when using these models for spatial extrapolation to
areas with the same relationship between predictors and the target
feature, but without any spatial correlation to the training data
itself. We defined the success of model evaluation methods as the
proportion of iterations which returned RMSE estimates between the 5th
and 9th percentile RMSEs of this ``ideal'' estimation procedure.

To find the error rate of the resubstitution approach, we fit 100 random
forests, one to each landscape, and then calculated the RMSE for each
model when used to predict its own training data. To find the error of
each CV approach, we first used each CV approach to separate each
landscape into \(n\) folds (Section~\ref{sec-resampling}). We then fit
models to each combination of \(n - 1\) of these folds, and calculated
RMSE when using the model to predict the remaining
\(D_{\operatorname{out}}\) (Equation~\ref{eq-rmse}).

\begin{equation}\protect\hypertarget{eq-rmse}{}{
\operatorname{RMSE} = \sqrt{(\frac{1}{n})\sum_{i=1}^{n}(y_{i} - \hat{y_{i}})^{2}}
}\label{eq-rmse}\end{equation}

We then calculated the variance of the RMSE estimates of each method
across the 100 simulated landscapes, as well as the proportion of runs
for each method which fell between the 5th and 95th percentiles of the
``true'' RMSE range.

Based upon prior research, we expected the optimal spacing between
\(D_{\operatorname{in}}\) and \(D_{\operatorname{out}}\) to be related
to the range of spatial dependence either in the outcome variable or in
model residuals (Le Rest et al. 2014; Roberts et al. 2017; Telford and
Birks 2009). As such, we quantified the range of spatial autocorrelation
in both the target variable \(y\) and in resubstitution residuals from
random forest models using the automated variogram fitting approach
implemented in the \texttt{automap} R package (Hiemstra et al. 2008).

\hypertarget{sec-results}{%
\section{Results and Discussion}\label{sec-results}}

\hypertarget{spatial-cv-improves-model-performance-estimates}{%
\subsection{Spatial CV Improves Model Performance
Estimates}\label{spatial-cv-improves-model-performance-estimates}}

\begin{figure}

{\centering \includegraphics{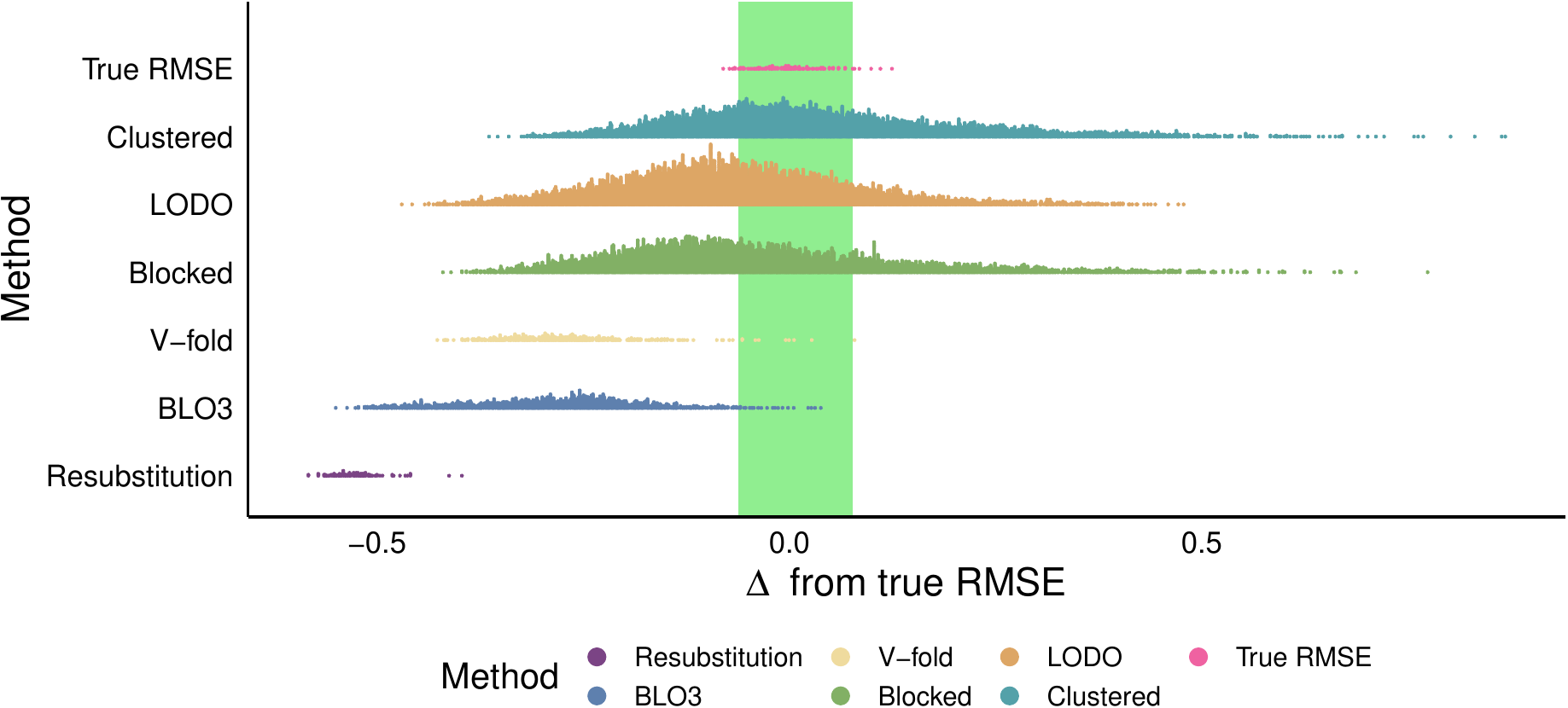}

}

\caption{\label{fig-comparisons}RMSE distributions of each spatial CV
method evaluated. The x axis represents the distance of an RMSE from the
mean ``true'' RMSE values, such that negative values underpredict the
true RMSE and positive values overpredict it. Distributions are scaled
relative to the number of iterations run, such that a unit area
represents the same number of iterations across each method, but not
necessarily the same proportion of all iterations evaluated. The green
rectangle represents the 90\% interval of true RMSE values used as the
``target'' RMSE range.}

\end{figure}

\begin{figure}

{\centering \includegraphics{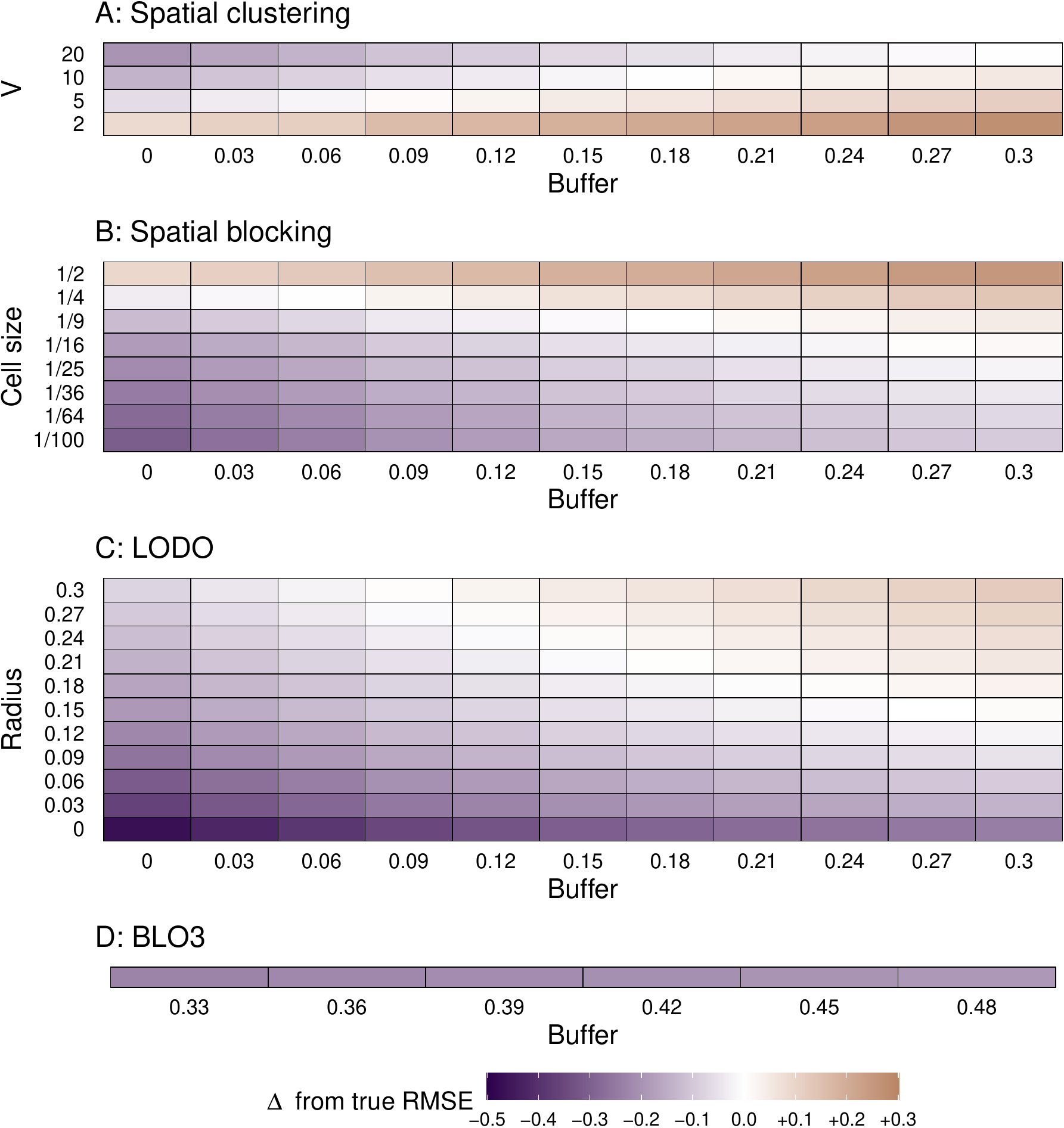}

}

\caption{\label{fig-rmse-delta}Spatial CV RMSE estimates under various
parameterizations. Colors represent the distance of an RMSE from the
mean ``true'' RMSE values, such that negative values underpredict the
true RMSE and positive values overpredict it. A: Spatial clustering RMSE
estimates using different numbers of clusters (``V'') and different
sizes of exclusion buffers (``Buffer''). B: Spatial blocking RMSE
estimates using different sizes of blocks (``Cell Size''; a cell size of
``1/2'' implies two blocks, each containing half the study area) and
different sizes of exclusion buffers (``Buffer''). C: LODO RMSE
estimates using different sizes of inclusion radii (``Radius'') and
different sizes of exclusion buffers (``Buffer''). D: BLO3 RMSE
estimates using different sizes of exclusion buffers (``Buffer'').}

\end{figure}

Spatial cross-validation methods consistently produced more accurate
estimates of model performance than non-spatial methods, which were
optimistically biased (producing too-low estimates of RMSE)
(Table~\ref{tbl-overall}; Figure~\ref{fig-comparisons}). CV produced the
best estimates when \(D_{\operatorname{out}}\) of spatially conjunct
observations were combined with exclusion buffers
(Table~\ref{tbl-winners}). Spatially clustered CV and LODO, both of
which enforce \(D_{\operatorname{out}}\) of spatially conjunct
observations, were among the most consistently effective CV methods
(Figure~\ref{fig-comparisons}). Removing too much data from
\(D_{\operatorname{in}}\), such as by clustering with only two folds or
blocking with only two blocks resulted in pessimistic over-estimates of
RMSE (Figure~\ref{fig-rmse-delta}).

\hypertarget{tbl-overall}{}
\begin{table}
\caption{\label{tbl-overall}Mean RMSE estimates across all evaluated parameterizations of
cross-validation strategies. Numbers in parentheses represent standard
deviations. ``\% within target RMSE range'' refers to the percentage of
iterations which had RMSE estimates between the 5th and 95th percentile
estimates from the true RMSE estimation procedure. }\tabularnewline

\centering
\begin{tabular}[t]{ll>{\raggedright\arraybackslash}p{7em}}
\toprule
Method & RMSE & \% within target RMSE range\\
\midrule
Ideal RMSE & 0.715 (0.042) & 90.00\%\\
\addlinespace
Clustered & 0.743 (0.161) & 36.97\%\\
\addlinespace
LODO & 0.641 (0.135) & 31.70\%\\
\addlinespace
Blocked & 0.664 (0.159) & 27.90\%\\
\addlinespace
V-fold & 0.440 (0.076) & 2.00\%\\
\addlinespace
BLO3CV & 0.429 (0.098) & 1.29\%\\
\addlinespace
Resubstitution & 0.189 (0.032) & 0.00\%\\
\bottomrule
\end{tabular}
\end{table}

\hypertarget{tbl-winners}{}
\begin{table}
\caption{\label{tbl-winners}Mean RMSE estimates across 100 iterations of various cross-validation
strategies (bold headers) for the parameterizations producing the most
accurate model performance estimates. Numbers in parentheses represent
standard deviations. ``\% within target RMSE range'' refers to the
percentage of iterations which had RMSE estimates between the 5th and
95th percentile estimates from the true RMSE estimation procedure. }\tabularnewline

\centering
\begin{tabular}[t]{rllrrl>{\raggedright\arraybackslash}p{7em}}
\toprule
V & Cell size & Cluster function & Exclusion buffer & Inclusion radius & RMSE & \% within target RMSE range\\
\midrule
\addlinespace[0.3em]
\multicolumn{7}{l}{\textbf{Ideal RMSE}}\\
\hspace{1em} &  &  &  &  & 0.715 (0.042) & 90.00\%\\
\addlinespace[0.3em]
\hline
\multicolumn{7}{l}{\textbf{Clustered}}\\
\hspace{1em}10 &  & kmeans & 0.15 &  & 0.694 (0.087) & 60.00\%\\
\hspace{1em}5 &  & kmeans & 0.09 &  & 0.723 (0.099) & 59.00\%\\
\hspace{1em}10 &  & kmeans & 0.18 &  & 0.712 (0.094) & 59.00\%\\
\addlinespace[0.3em]
\hline
\multicolumn{7}{l}{\textbf{LODO}}\\
\hspace{1em} &  &  & 0.18 & 0.21 & 0.718 (0.095) & 60.00\%\\
\hspace{1em} &  &  & 0.12 & 0.24 & 0.703 (0.093) & 59.00\%\\
\hspace{1em} &  &  & 0.12 & 0.27 & 0.725 (0.098) & 59.00\%\\
\addlinespace[0.3em]
\hline
\multicolumn{7}{l}{\textbf{Blocked}}\\
\hspace{1em} & 1/9 &  & 0.24 &  & 0.738 (0.099) & 61.00\%\\
\hspace{1em} & 1/9 &  & 0.21 &  & 0.732 (0.100) & 60.00\%\\
\hspace{1em} & 1/25 &  & 0.27 &  & 0.688 (0.084) & 58.00\%\\
\addlinespace[0.3em]
\hline
\multicolumn{7}{l}{\textbf{V-fold}}\\
\hspace{1em}2 &  &  &  &  & 0.475 (0.079) & 2.00\%\\
\hspace{1em}5 &  &  &  &  & 0.438 (0.073) & 2.00\%\\
\hspace{1em}10 &  &  &  &  & 0.428 (0.071) & 2.00\%\\
\addlinespace[0.3em]
\hline
\multicolumn{7}{l}{\textbf{BLO3}}\\
\hspace{1em} &  &  & 0.48 &  & 0.524 (0.070) & 7.00\%\\
\hspace{1em} &  &  & 0.45 &  & 0.516 (0.069) & 4.00\%\\
\hspace{1em} &  &  & 0.42 &  & 0.508 (0.067) & 3.00\%\\
\addlinespace[0.3em]
\hline
\multicolumn{7}{l}{\textbf{Resubstitution}}\\
\hspace{1em} &  &  &  &  & 0.189 (0.032) & 0.00\%\\
\bottomrule
\end{tabular}
\end{table}

\begin{figure}

{\centering \includegraphics{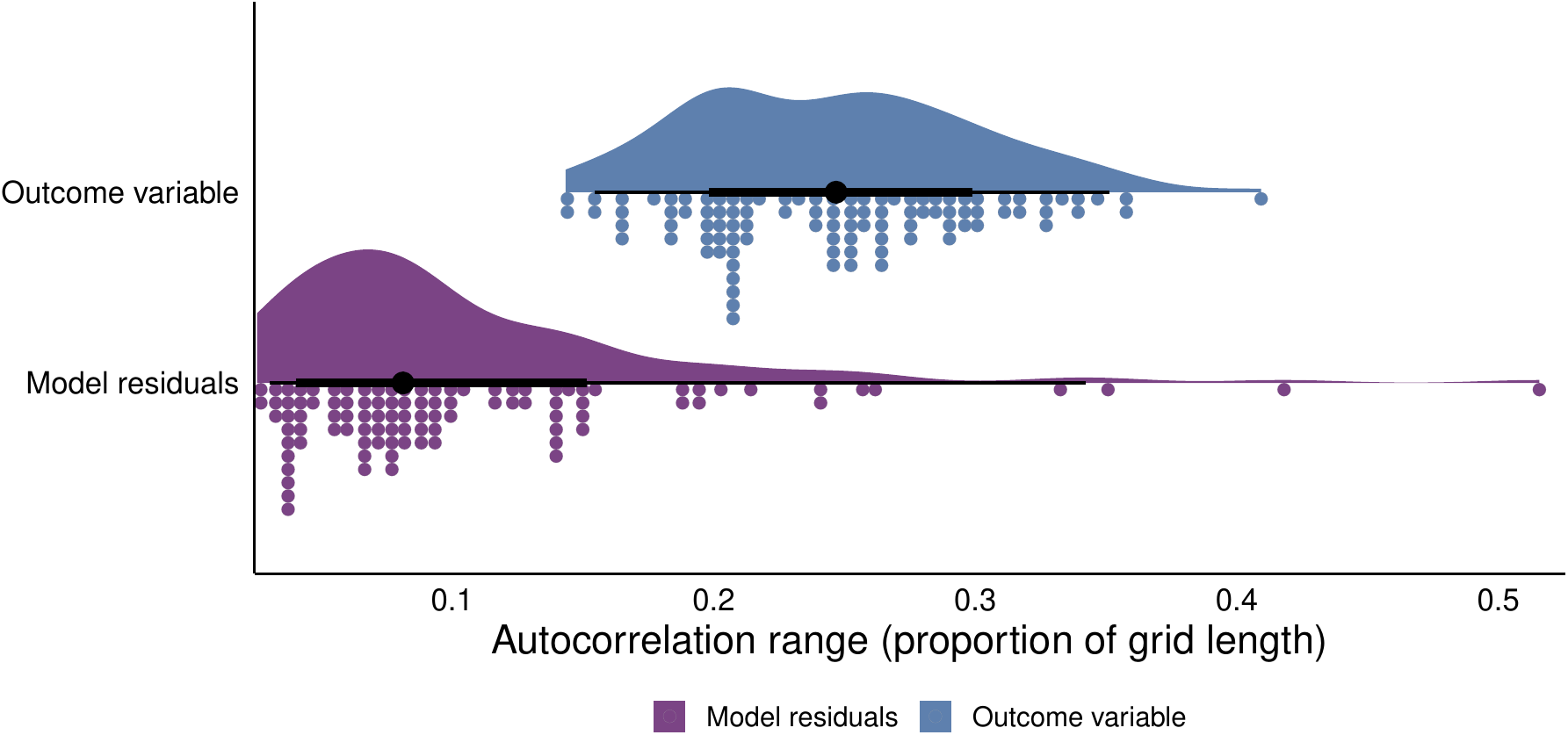}

}

\caption{\label{fig-ranges}Autocorrelation range distributions for the
outcome variable and resubstitution model residuals, used throughout the
literature to identify target distances to separate
\(D_{\operatorname{out}}\) and \(D_{\operatorname{in}}\). Units are as a
proportion of the side length of the grid. Ranges were determined via
empirical variogram. Each point represents one iteration of the
simulation process, while ``clouds'' represent the matching probability
density function. Black points represent the median autocorrelation
range, while the black bar represents the interquartile range and 95\%
interval.}

\end{figure}

The best parameter sets for CV methods consistently separated the center
of \(D_{\operatorname{out}}\) from \(D_{\operatorname{in}}\) by
25\%-41\% of the grid length (Clustering 25\%-29\%; LODO 36\%-39\%;
Blocked 32\%-41\%; BLO3 24\%-30\%). Given that the target variable had a
mean autocorrelation range of 24.61\% of the grid length
(Figure~\ref{fig-ranges}), this suggests that spatial cross-validation
approaches produce the best estimates of model performance when
\(D_{\operatorname{out}}\) is sufficiently separated from
\(D_{\operatorname{in}}\) such that there is no spatial dependency in
the outcome variable between the two sets.

\begin{figure}

{\centering \includegraphics{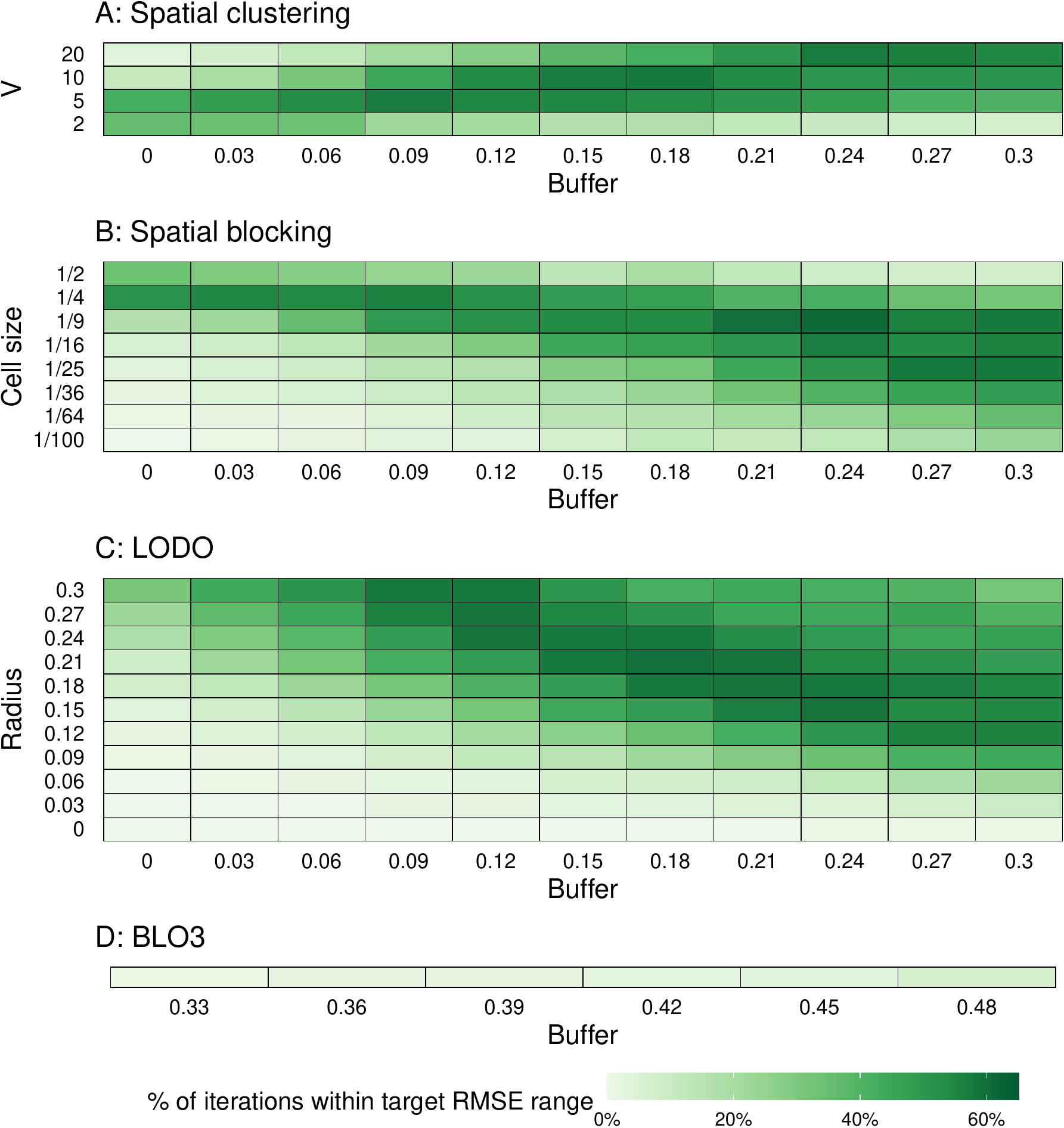}

}

\caption{\label{fig-rmse-prop}Percentage of RMSE estimates within the
target range (the 90\% interval of true RMSE values) from spatial CV
methods under various parameterizations. A: Spatial clustering using
different numbers of clusters (``V'') and different sizes of exclusion
buffers (``Buffer''). B: Spatial blocking using different sizes of
blocks (``Cell Size''; a cell size of ``1/2'' implies two blocks, each
containing half the study area) and different sizes of exclusion buffers
(``Buffer''). C: LODO using different sizes of inclusion radii
(``Radius'') and different sizes of exclusion buffers (``Buffer''). D:
BLO3 using different sizes of exclusion buffers (``Buffer'').}

\end{figure}

Clustering appeared to be the spatial CV method most robust to different
parameterizations (Figure~\ref{fig-comparisons};
Figure~\ref{fig-rmse-delta}), with the highest proportion of all
iterations within the target RMSE range (Clustered 36.97\% of all
iterations; LODO 31.70\%; Blocked 27.90\%; BLO3 )
(Figure~\ref{fig-rmse-prop}). This may, however, simply reflect the
relatively narrow range of parameters evaluated with clustering, as both
blocking and LODO had a wide range of parameters which returned
estimates within the target range at least half the time
(Figure~\ref{fig-rmse-prop}). While BLO3 exhibited increasing RMSE with
increasing buffer radii (Figure~\ref{fig-rmse-delta}), as frequently
reported in the literature, we found it only rarely produced RMSE
estimates within the target range (Figure~\ref{fig-rmse-prop}).

\begin{figure}

{\centering \includegraphics{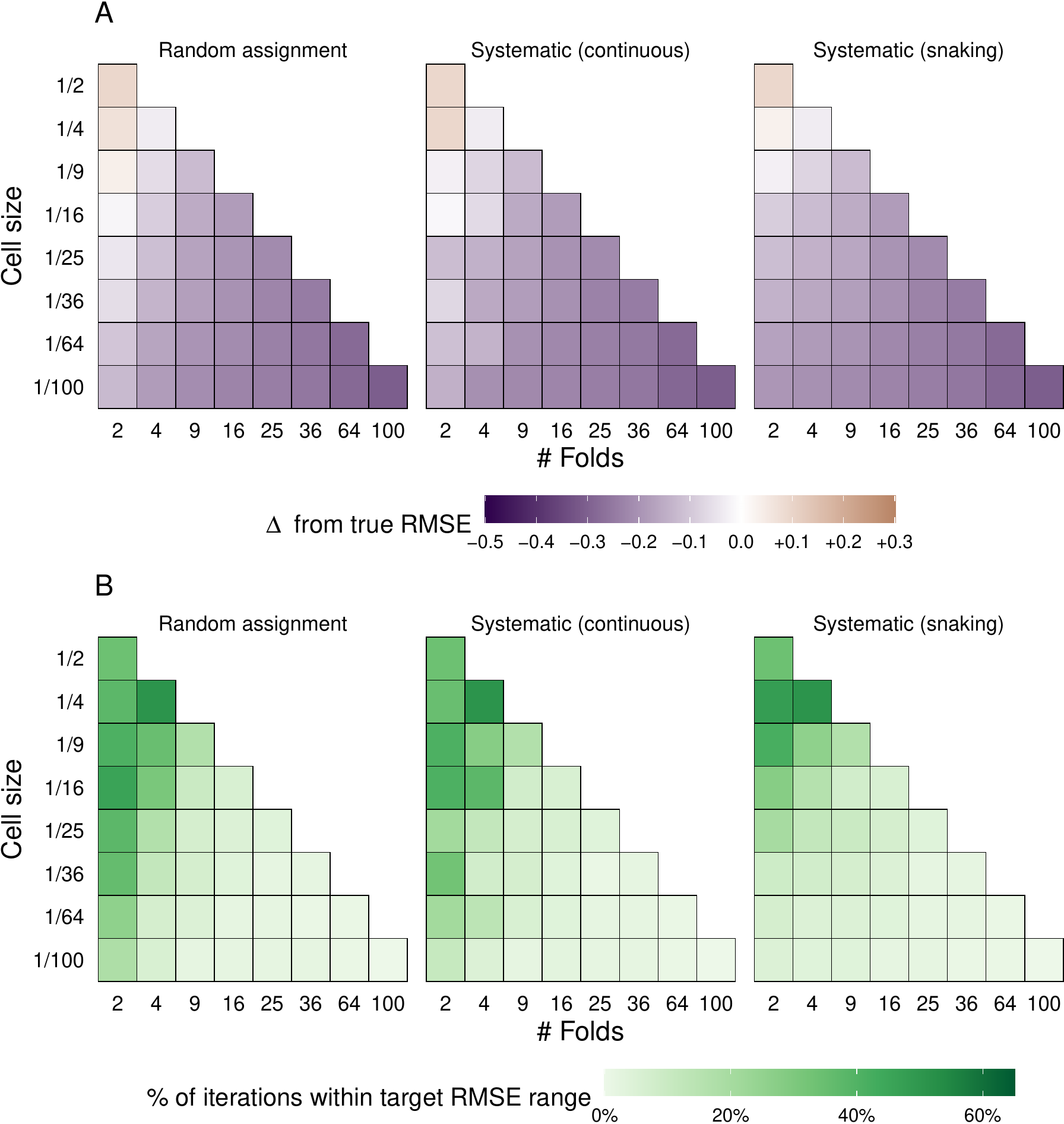}

}

\caption{\label{fig-blocks-with-v}RMSE estimates from spatially blocked
CV with various block sizes assigned to various numbers of
\(D_{\operatorname{out}}\). A: Spatial CV RMSE estimates under various
parameterizations. Colors represent the distance of an RMSE from the
mean ``true'' RMSE values, such that negative values underpredict the
true RMSE and positive values overpredict it. B: Percentage of RMSE
estimates within the target range (the 90\% interval of true RMSE
values). Labels above each panel refer to the method for assigning
blocks to folds: either random assignment, systematically with each row
assigned from left to right (``continuous''), or systematically with
each row assigned in a ``snaking'' pattern (first row left to right,
next row right to left, then repeating).}

\end{figure}

RMSE estimates from spatial blocking were inversely related to the
number of \(D_{\operatorname{out}}\) used, likely due to the distance
between \(D_{\operatorname{out}}\) and \(D_{\operatorname{in}}\)
increasing if adjacent blocks were assigned to the same
\(D_{\operatorname{out}}\) (Figure~\ref{fig-blocks-with-v}). RMSE
estimates generally increased gradually as the number of
\(D_{\operatorname{out}}\) decreased, though notable increases were
observed when blocks were assigned via the continuous systematic method
and the number of cells in each grid row were evenly divisible by the
number of \(D_{\operatorname{out}}\) (e.g., when 1/16th cell sizes
produced by a 4-by-4 grid were divided into 4 folds). In these
situations, each column of the grid will be entirely assigned to the
same \(D_{\operatorname{out}}\), somewhat resembling the CV method of
Wenger and Olden (2012), producing \(D_{\operatorname{out}}\) which have
no neighboring \(D_{\operatorname{in}}\) observations in the \emph{y}
direction and therefore a greater average distance between
\(D_{\operatorname{out}}\) and \(D_{\operatorname{in}}\). Overall, these
results suggest that using fewer \(D_{\operatorname{out}}\) than blocks
may be appropriate when the range of autocorrelation in the outcome
variable is relatively small and there is concern about large blocks
restricting predictor space (Roberts et al. 2017); however, with longer
autocorrelation ranges it is likely best to use a leave-one-block-out
approach with fewer, larger blocks.

As such, the recommendations from this study are clear: CV-based
performance assessments of models fit using spatial data benefit from
spatial CV approaches. Those spatial CV approaches are most likely to
return good estimates of true model accuracy if they combine
\(D_{\operatorname{out}}\) of spatially conjunct observations with
exclusion buffers, such that the average observation is separated from
\(D_{\operatorname{in}}\) by enough distance that there is no spatial
dependency in the outcome variable between \(D_{\operatorname{in}}\) and
\(D_{\operatorname{out}}\).

\hypertarget{limitations}{%
\subsection{Limitations}\label{limitations}}

This simulation study assumed that spatial CV could take advantage of
regularly distributed observations, such that all locations had a
similar density of measurement points. This assumption is often
violated, as it is often impractical to obtain a uniform sample across
large areas, and as such observations are often clustered in more
convenient locations and relatively sparse in less accessible areas
(Meyer and Pebesma 2022; Martin, Blossey, and Ellis 2012). Alternative
approaches not investigated in this study may be more effective in these
situations; for instance, when the expected distance between training
data and model predictions is known, Milà et al. (2022) proposes an
alternative nearest neighbor distance matching CV approach which may
equal or improve upon buffered leave-one-out CV. An alternative approach
put forward by Meyer et al. (2018) uses meaningful, human-defined
locations as groups for CV, which may produce better results than the
automated partitioning methods investigated in this study. While we
believe our results clearly demonstrate the benefits of spatial CV for
sampling designs resembling our simulation, we do not pretend to present
the one CV approach to rule them all.

We additionally do not expect these results, focused upon using CV to
evaluate the accuracy of predictive models, will necessarily transfer to
map accuracy assessments. Stehman and Foody (2019) explained that
design-based sampling approaches provide unbiased assessments of map
accuracy, while Wadoux et al. (2021) demonstrated that spatial CV
methods may be overly pessimistic when assessing map accuracy, and Bruin
et al. (2022) suggested sampling-intensity weighted CV approaches for
map accuracy assessments in situations where the study area has been
unevenly sampled. However, we expect these results will be informative
in the many situations requiring estimates of model accuracy,
particularly given that traditional held-out test sets are somewhat rare
in the spatial modeling literature.

Lastly, we did not investigate any CV approaches which aim to preserve
outcome or predictor distributions across \(D_{\operatorname{out}}\).
When working with imbalanced outcomes, random sampling may produce
\(D_{\operatorname{out}}\) with notably different outcome distributions
than the overall training data, which may bias performance estimates.
Assigning stratified samples of observations to
\(D_{\operatorname{out}}\) can address this, but it is not obvious how
to use stratified sampling when assigning groups of observations (such
as a spatial cluster or block), with one outcome value per observation,
to \(D_{\operatorname{out}}\) as a unit. The \texttt{rsample} package
allows stratified CV when all observations within a given group have
identical outcome values (that is, when groups are strictly nested
within the stratification variable), but this condition is rare and
difficult to enforce when using unsupervised group assignment based on
spatial location, as with all the spatial CV methods we investigated.

Creating \(D_{\operatorname{out}}\) based on predictor space, rather
than outcome distributions, has also been proposed as a solution to
spatial CV procedures restricting the predictor ranges present in
\(D_{\operatorname{in}}\). This is a particular challenge if the
predictors themselves are spatially structured, and may unintentionally
force models to extrapolate further in predictor space than would be
expected when predicting new data (Roberts et al. 2017). As increasing
distance in predictor space often correlates with increasing error (e.g.
Thuiller et al. 2004; Sheridan et al. 2004; Meyer and Pebesma 2021),
Roberts et al. (2017) suggest blocking approaches to minimize distance
in predictor space between folds, although to the best of our knowledge
these approaches are not yet in widespread use. A related field of
research suggests methods for calculating the applicability domain of a
model (Netzeva et al. 2005; Meyer and Pebesma 2021), which can help to
identify when predicting new observations will require extrapolation in
predictor space, and will likely produce predictions with higher than
expected errors. Such methods are particularly well-equipped to
supplement spatial CV procedures, as it adjusts the permissible distance
in predictor space based upon the distance between
\(D_{\operatorname{in}}\) and \(D_{\operatorname{out}}\).

\hypertarget{sec-conclusion}{%
\section{Conclusion}\label{sec-conclusion}}

These results reinforce that spatial CV is essential for evaluating the
performance of predictive models fit to data with internal spatial
structure, particularly in situations where design-based map accuracy
assessments are not practical or germane. Techniques that apply
exclusion buffers around assessment sets of spatially conjunct
observations, such as spatial clustering and LODO, are likely to produce
the best estimates of model performance. The most accurate estimates of
model performance are produced when the assessment and analysis data are
sufficiently separated so that there is no spatial dependence in the
outcome variable between the two sets.

\hypertarget{acknowledgements}{%
\section{Acknowledgements}\label{acknowledgements}}

We would like to thank Posit, PBC, for support in the development of the
\texttt{rsample} and \texttt{spatialsample} packages.

\hypertarget{software-data-and-code-availability}{%
\section{Software, data, and code
availability}\label{software-data-and-code-availability}}

The \texttt{spatialsample} package is available online at
https://github.com/tidymodels/spatialsample . All data and code used in
this paper are available online at
https://github.com/cafri-labs/assessing-spatial-cv .

\newpage{}

\hypertarget{references}{%
\section*{References}\label{references}}
\addcontentsline{toc}{section}{References}

\hypertarget{refs}{}
\begin{CSLReferences}{1}{0}
\leavevmode\vadjust pre{\hypertarget{ref-adams2020}{}}%
Adams, Matthew D., Felix Massey, Karl Chastko, and Calvin Cupini. 2020.
{``Spatial Modelling of Particulate Matter Air Pollution Sensor
Measurements Collected by Community Scientists While Cycling, Land Use
Regression with Spatial Cross-Validation, and Applications of Machine
Learning for Data Correction.''} \emph{Atmospheric Environment} 230
(June): 117479. \url{https://doi.org/10.1016/j.atmosenv.2020.117479}.

\leavevmode\vadjust pre{\hypertarget{ref-bahn2012}{}}%
Bahn, Volker, and Brian J. McGill. 2012. {``Testing the Predictive
Performance of Distribution Models.''} \emph{Oikos} 122 (3): 321--31.
\url{https://doi.org/10.1111/j.1600-0706.2012.00299.x}.

\leavevmode\vadjust pre{\hypertarget{ref-bastin2019}{}}%
Bastin, Jean-Francois, Yelena Finegold, Claude Garcia, Danilo Mollicone,
Marcelo Rezende, Devin Routh, Constantin M. Zohner, and Thomas W.
Crowther. 2019. {``The Global Tree Restoration Potential.''}
\emph{Science} 365 (6448): 76--79.
\url{https://doi.org/10.1126/science.aax0848}.

\leavevmode\vadjust pre{\hypertarget{ref-Bates2021}{}}%
Bates, Stephen, Trevor Hastie, and Robert Tibshirani. 2021.
{``Cross-Validation: What Does It Estimate and How Well Does It Do It?
arXiv:2104.00673v2 {[}Stat.ME{]}.''}
\url{https://doi.org/10.48550/arXiv.2104.00673}.

\leavevmode\vadjust pre{\hypertarget{ref-Breiman2001}{}}%
Breiman, Leo. 2001. {``{Random Forests}.''} \emph{Machine Learning} 45:
5--32. \url{https://doi.org/10.1023/A:1010933404324}.

\leavevmode\vadjust pre{\hypertarget{ref-brenning2012}{}}%
Brenning, Alexander. 2012. {``Spatial Cross-Validation and Bootstrap for
the Assessment of Prediction Rules in Remote Sensing: The {R} Package
{sperrorest}.''} \emph{2012 IEEE International Geoscience and Remote
Sensing Symposium}, July.
\url{https://doi.org/10.1109/igarss.2012.6352393}.

\leavevmode\vadjust pre{\hypertarget{ref-debruin2022}{}}%
Bruin, Sytze de, Dick J. Brus, Gerard B. M. Heuvelink, Tom van
Ebbenhorst Tengbergen, and Alexandre M.J-C. Wadoux. 2022. {``Dealing
with Clustered Samples for Assessing Map Accuracy by
Cross-Validation.''} \emph{Ecological Informatics} 69 (July): 101665.
\url{https://doi.org/10.1016/j.ecoinf.2022.101665}.

\leavevmode\vadjust pre{\hypertarget{ref-brus2020}{}}%
Brus, Dick J. 2020. {``Statistical Approaches for Spatial Sample Survey:
Persistent Misconceptions and New Developments.''} \emph{European
Journal of Soil Science} 72 (2): 686--703.
\url{https://doi.org/10.1111/ejss.12988}.

\leavevmode\vadjust pre{\hypertarget{ref-s2}{}}%
Dunnington, Dewey, Edzer Pebesma, and Ege Rubak. 2021. \emph{{s2}:
Spherical Geometry Operators Using the {S2} Geometry Library}.
\url{https://CRAN.R-project.org/package=s2}.

\leavevmode\vadjust pre{\hypertarget{ref-efron1986}{}}%
Efron, Bradley. 1986. {``How Biased Is the Apparent Error Rate of a
Prediction Rule?''} \emph{Journal of the American Statistical
Association} 81 (394): 461--70.
\url{https://doi.org/10.1080/01621459.1986.10478291}.

\leavevmode\vadjust pre{\hypertarget{ref-efron1983}{}}%
Efron, Bradley, and Gail Gong. 1983. {``A Leisurely Look at the
Bootstrap, the Jackknife, and Cross-Validation.''} \emph{The American
Statistician} 37 (1): 36--48.
\url{https://doi.org/10.1080/00031305.1983.10483087}.

\leavevmode\vadjust pre{\hypertarget{ref-fick2017}{}}%
Fick, Stephen E., and Robert J. Hijmans. 2017. {``WorldClim 2: New
1{-}Km Spatial Resolution Climate Surfaces for Global Land Areas.''}
\emph{International Journal of Climatology} 37 (12): 4302--15.
\url{https://doi.org/10.1002/joc.5086}.

\leavevmode\vadjust pre{\hypertarget{ref-rsample}{}}%
Frick, Hannah, Fanny Chow, Max Kuhn, Michael Mahoney, Julia Silge, and
Hadley Wickham. 2022. \emph{{rsample}: General Resampling
Infrastructure}. \url{https://CRAN.R-project.org/package=rsample}.

\leavevmode\vadjust pre{\hypertarget{ref-gong1986}{}}%
Gong, Gail. 1986. {``Cross-Validation, the Jackknife, and the Bootstrap:
Excess Error Estimation in Forward Logistic Regression.''} \emph{Journal
of the American Statistical Association} 81 (393): 108--13.
\url{https://doi.org/10.1080/01621459.1986.10478245}.

\leavevmode\vadjust pre{\hypertarget{ref-degruijter1990}{}}%
Gruijter, J. J. de, and C. J. F. ter Braak. 1990. {``Model-Free
Estimation from Spatial Samples: A Reappraisal of Classical Sampling
Theory.''} \emph{Mathematical Geology} 22 (4): 407--15.
\url{https://doi.org/10.1007/bf00890327}.

\leavevmode\vadjust pre{\hypertarget{ref-hengl2017}{}}%
Hengl, Tomislav, Jorge Mendes de Jesus, Gerard B. M. Heuvelink, Maria
Ruiperez Gonzalez, Milan Kilibarda, Aleksandar Blagotić, Wei Shangguan,
et al. 2017. {``SoilGrids250m: Global Gridded Soil Information Based on
Machine Learning.''} Edited by Ben Bond-Lamberty. \emph{PLOS ONE} 12
(2): e0169748. \url{https://doi.org/10.1371/journal.pone.0169748}.

\leavevmode\vadjust pre{\hypertarget{ref-automap}{}}%
Hiemstra, P. H., E. J. Pebesma, C. J. W. Twenhöfel, and G. B. M.
Heuvelink. 2008. {``Real-Time Automatic Interpolation of Ambient Gamma
Dose Rates from the Dutch Radioactivity Monitoring Network.''}
\emph{Computers \& Geosciences}.
\url{https://doi.org/10.1016/j.cageo.2008.10.011}.

\leavevmode\vadjust pre{\hypertarget{ref-vandenhoogen2019}{}}%
Hoogen, Johan van den, Stefan Geisen, Devin Routh, Howard Ferris, Walter
Traunspurger, David A. Wardle, Ron G. M. de Goede, et al. 2019. {``Soil
Nematode Abundance and Functional Group Composition at a Global
Scale.''} \emph{Nature} 572 (7768): 194--98.
\url{https://doi.org/10.1038/s41586-019-1418-6}.

\leavevmode\vadjust pre{\hypertarget{ref-karasiak2021}{}}%
Karasiak, N., J.-F. Dejoux, C. Monteil, and D. Sheeren. 2021. {``Spatial
Dependence Between Training and Test Sets: Another Pitfall of
Classification Accuracy Assessment in Remote Sensing.''} \emph{Machine
Learning} 111 (7): 2715--40.
\url{https://doi.org/10.1007/s10994-021-05972-1}.

\leavevmode\vadjust pre{\hypertarget{ref-tune}{}}%
Kuhn, Max. 2022. \emph{{tune}: Tidy Tuning Tools}.
\url{https://CRAN.R-project.org/package=tune}.

\leavevmode\vadjust pre{\hypertarget{ref-dials}{}}%
Kuhn, Max, and Hannah Frick. 2022. \emph{{dials}: Tools for Creating
Tuning Parameter Values}.
\url{https://CRAN.R-project.org/package=dials}.

\leavevmode\vadjust pre{\hypertarget{ref-Kuhn2013}{}}%
Kuhn, Max, and Kjell Johnson. 2013. \emph{Applied Predictive Modeling}.
Vol. 26. Springer. \url{https://doi.org/10.1007/978-1-4614-6849-3}.

\leavevmode\vadjust pre{\hypertarget{ref-kuhn2019}{}}%
---------. 2019. \emph{Feature Engineering and Selection}. Chapman;
Hall/CRC. \url{https://doi.org/10.1201/9781315108230}.

\leavevmode\vadjust pre{\hypertarget{ref-tmwr}{}}%
Kuhn, Max, and Julia Silge. 2022. \emph{Tidy Modeling with {R}}.
O'Reilly.

\leavevmode\vadjust pre{\hypertarget{ref-lerest2014}{}}%
Le Rest, Kévin, David Pinaud, Pascal Monestiez, Joël Chadoeuf, and
Vincent Bretagnolle. 2014. {``Spatial Leave-One-Out Cross-Validation for
Variable Selection in the Presence of Spatial Autocorrelation.''}
\emph{Global Ecology and Biogeography} 23 (7): 811--20.
\url{https://doi.org/10.1111/geb.12161}.

\leavevmode\vadjust pre{\hypertarget{ref-legendre1989}{}}%
Legendre, Pierre, and Marie Josée Fortin. 1989. {``Spatial Pattern and
Ecological Analysis.''} \emph{Vegetatio} 80 (2): 107--38.
\url{https://doi.org/10.1007/bf00048036}.

\leavevmode\vadjust pre{\hypertarget{ref-spatialsample}{}}%
Mahoney, Michael, and Julia Silge. 2022. \emph{{spatialsample}: Spatial
Resampling Infrastructure}.
\url{https://CRAN.R-project.org/package=spatialsample}.

\leavevmode\vadjust pre{\hypertarget{ref-martin2012}{}}%
Martin, Laura J, Bernd Blossey, and Erle Ellis. 2012. {``Mapping Where
Ecologists Work: Biases in the Global Distribution of Terrestrial
Ecological Observations.''} \emph{Frontiers in Ecology and the
Environment} 10 (4): 195--201. \url{https://doi.org/10.1890/110154}.

\leavevmode\vadjust pre{\hypertarget{ref-meyer2021}{}}%
Meyer, Hanna, and Edzer Pebesma. 2021. {``Predicting into Unknown Space?
Estimating the Area of Applicability of Spatial Prediction Models.''}
\emph{Methods in Ecology and Evolution} 12 (9): 1620--33.
\url{https://doi.org/10.1111/2041-210x.13650}.

\leavevmode\vadjust pre{\hypertarget{ref-meyer2022}{}}%
---------. 2022. {``Machine Learning-Based Global Maps of Ecological
Variables and the Challenge of Assessing Them.''} \emph{Nature
Communications} 13 (1).
\url{https://doi.org/10.1038/s41467-022-29838-9}.

\leavevmode\vadjust pre{\hypertarget{ref-meyer2018}{}}%
Meyer, Hanna, Christoph Reudenbach, Tomislav Hengl, Marwan Katurji, and
Thomas Nauss. 2018. {``Improving Performance of Spatio-Temporal Machine
Learning Models Using Forward Feature Selection and Target-Oriented
Validation.''} \emph{Environmental Modelling \& Software} 101 (March):
1--9. \url{https://doi.org/10.1016/j.envsoft.2017.12.001}.

\leavevmode\vadjust pre{\hypertarget{ref-meyer2019}{}}%
Meyer, Hanna, Christoph Reudenbach, Stephan Wöllauer, and Thomas Nauss.
2019. {``Importance of Spatial Predictor Variable Selection in Machine
Learning Applications {\textendash} Moving from Data Reproduction to
Spatial Prediction.''} \emph{Ecological Modelling} 411 (November):
108815. \url{https://doi.org/10.1016/j.ecolmodel.2019.108815}.

\leavevmode\vadjust pre{\hypertarget{ref-miluxe02022}{}}%
Milà, Carles, Jorge Mateu, Edzer Pebesma, and Hanna Meyer. 2022.
{``Nearest Neighbour Distance Matching Leave{-}One{-}Out
Cross{-}Validation for Map Validation.''} \emph{Methods in Ecology and
Evolution} 13 (6): 1304--16.
\url{https://doi.org/10.1111/2041-210x.13851}.

\leavevmode\vadjust pre{\hypertarget{ref-netzeva2005}{}}%
Netzeva, Tatiana I., Andrew P. Worth, Tom Aldenberg, Romualdo Benigni,
Mark T. D. Cronin, Paola Gramatica, Joanna S. Jaworska, et al. 2005.
{``Current Status of Methods for Defining the Applicability Domain of
(Quantitative) Structure-Activity Relationships.''} \emph{Alternatives
to Laboratory Animals} 33 (2): 155--73.
\url{https://doi.org/10.1177/026119290503300209}.

\leavevmode\vadjust pre{\hypertarget{ref-osullivan2010}{}}%
O'Sullivan, David, and David J. Unwin. 2010. \emph{Geographic
Information Analysis}. John Wiley \& Sons, Inc.
\url{https://doi.org/10.1002/9780470549094}.

\leavevmode\vadjust pre{\hypertarget{ref-sf}{}}%
Pebesma, Edzer. 2018. {``{Simple Features for R: Standardized Support
for Spatial Vector Data}.''} \emph{{The R Journal}} 10 (1): 439--46.
\url{https://doi.org/10.32614/RJ-2018-009}.

\leavevmode\vadjust pre{\hypertarget{ref-units}{}}%
Pebesma, Edzer, Thomas Mailund, and James Hiebert. 2016. {``Measurement
Units in {R}.''} \emph{R Journal} 8 (2): 486--94.
\url{https://doi.org/10.32614/RJ-2016-061}.

\leavevmode\vadjust pre{\hypertarget{ref-ploton2020}{}}%
Ploton, Pierre, Frédéric Mortier, Maxime Réjou-Méchain, Nicolas Barbier,
Nicolas Picard, Vivien Rossi, Carsten Dormann, et al. 2020. {``Spatial
Validation Reveals Poor Predictive Performance of Large-Scale Ecological
Mapping Models.''} \emph{Nature Communications} 11 (1).
\url{https://doi.org/10.1038/s41467-020-18321-y}.

\leavevmode\vadjust pre{\hypertarget{ref-pohjankukka2017}{}}%
Pohjankukka, Jonne, Tapio Pahikkala, Paavo Nevalainen, and Jukka
Heikkonen. 2017. {``Estimating the Prediction Performance of Spatial
Models via Spatial k-Fold Cross Validation.''} \emph{International
Journal of Geographical Information Science} 31 (10): 2001--19.
\url{https://doi.org/10.1080/13658816.2017.1346255}.

\leavevmode\vadjust pre{\hypertarget{ref-Probst2018}{}}%
Probst, Philipp, Bernd Bischl, and Anne-Laure Boulesteix. 2018.
{``Tunability: Importance of Hyperparameters of Machine Learning
Algorithms.''} arXiv. \url{https://doi.org/10.48550/ARXIV.1802.09596}.

\leavevmode\vadjust pre{\hypertarget{ref-R}{}}%
R Core Team. 2022. \emph{{R}: A Language and Environment for Statistical
Computing}. Vienna, Austria: R Foundation for Statistical Computing.
\url{https://www.R-project.org/}.

\leavevmode\vadjust pre{\hypertarget{ref-Roberts2017}{}}%
Roberts, David R., Volker Bahn, Simone Ciuti, Mark S. Boyce, Jane Elith,
Gurutzeta Guillera-Arroita, Severin Hauenstein, et al. 2017.
{``Cross-Validation Strategies for Data with Temporal, Spatial,
Hierarchical, or Phylogenetic Structure.''} \emph{Ecography} 40 (8):
913--29. https://doi.org/\url{https://doi.org/10.1111/ecog.02881}.

\leavevmode\vadjust pre{\hypertarget{ref-RandomFields}{}}%
Schlather, Martin, Alexander Malinowski, Peter J. Menck, Marco Oesting,
and Kirstin Strokorb. 2015. {``Analysis, Simulation and Prediction of
Multivariate Random Fields with Package {RandomFields}.''} \emph{Journal
of Statistical Software} 63 (8): 1--25.
\url{https://doi.org/10.18637/jss.v063.i08}.

\leavevmode\vadjust pre{\hypertarget{ref-schratz2019}{}}%
Schratz, Patrick, Jannes Muenchow, Eugenia Iturritxa, Jakob Richter, and
Alexander Brenning. 2019. {``Hyperparameter Tuning and Performance
Assessment of Statistical and Machine-Learning Algorithms Using Spatial
Data.''} \emph{Ecological Modelling} 406 (August): 109--20.
\url{https://doi.org/10.1016/j.ecolmodel.2019.06.002}.

\leavevmode\vadjust pre{\hypertarget{ref-sheridan2004}{}}%
Sheridan, Robert P., Bradley P. Feuston, Vladimir N. Maiorov, and Simon
K. Kearsley. 2004. {``Similarity to Molecules in the Training Set Is a
Good Discriminator for Prediction Accuracy in {QSAR}.''} \emph{Journal
of Chemical Information and Computer Sciences} 44 (6): 1912--28.
\url{https://doi.org/10.1021/ci049782w}.

\leavevmode\vadjust pre{\hypertarget{ref-stehman2019}{}}%
Stehman, Stephen V., and Giles M. Foody. 2019. {``Key Issues in Rigorous
Accuracy Assessment of Land Cover Products.''} \emph{Remote Sensing of
Environment} 231 (September): 111199.
\url{https://doi.org/10.1016/j.rse.2019.05.018}.

\leavevmode\vadjust pre{\hypertarget{ref-Stone1974}{}}%
Stone, M. 1974. {``Cross-Validatory Choice and Assessment of Statistical
Predictions.''} \emph{Journal of the Royal Statistical Society. Series B
(Methodological)} 36 (2): 111--47.
\url{http://www.jstor.org/stable/2984809}.

\leavevmode\vadjust pre{\hypertarget{ref-telford2009}{}}%
Telford, R. J., and H. J. B. Birks. 2009. {``Evaluation of Transfer
Functions in Spatially Structured Environments.''} \emph{Quaternary
Science Reviews} 28 (13-14): 1309--16.
\url{https://doi.org/10.1016/j.quascirev.2008.12.020}.

\leavevmode\vadjust pre{\hypertarget{ref-thuiller2004}{}}%
Thuiller, Wilfried, Lluis Brotons, Miguel B. Araújo, and Sandra Lavorel.
2004. {``Effects of Restricting Environmental Range of Data to Project
Current and Future Species Distributions.''} \emph{Ecography} 27 (2):
165--72. \url{https://doi.org/10.1111/j.0906-7590.2004.03673.x}.

\leavevmode\vadjust pre{\hypertarget{ref-townsend2007}{}}%
Townsend, Peterson A., Monica Papeş, and Muir Eaton. 2007.
{``Transferability and Model Evaluation in Ecological Niche Modeling: A
Comparison of {GARP} and {Maxent}.''} \emph{Ecography} 30 (4): 550--60.
\url{https://doi.org/10.1111/j.0906-7590.2007.05102.x}.

\leavevmode\vadjust pre{\hypertarget{ref-valavi2018}{}}%
Valavi, Roozbeh, Jane Elith, José J. Lahoz-Monfort, and Gurutzeta
Guillera-Arroita. 2018. {``{blockCV}: An {R} Package for Generating
Spatially or Environmentally Separated Folds for k-Fold Cross-Validation
of Species Distribution Models.''} Edited by David Warton. \emph{Methods
in Ecology and Evolution} 10 (2): 225--32.
\url{https://doi.org/10.1111/2041-210x.13107}.

\leavevmode\vadjust pre{\hypertarget{ref-varma2006}{}}%
Varma, Sudhir, and Richard Simon. 2006. {``Bias in Error Estimation When
Using Cross-Validation for Model Selection.''} \emph{BMC Bioinformatics}
7 (1). \url{https://doi.org/10.1186/1471-2105-7-91}.

\leavevmode\vadjust pre{\hypertarget{ref-wadoux2021}{}}%
Wadoux, Alexandre M. J.-C., Gerard B. M. Heuvelink, Sytze de Bruin, and
Dick J. Brus. 2021. {``Spatial Cross-Validation Is Not the Right Way to
Evaluate Map Accuracy.''} \emph{Ecological Modelling} 457 (October):
109692. \url{https://doi.org/10.1016/j.ecolmodel.2021.109692}.

\leavevmode\vadjust pre{\hypertarget{ref-walvoort2010}{}}%
Walvoort, D. J. J., D. J. Brus, and J. J. de Gruijter. 2010. {``An {R}
Package for Spatial Coverage Sampling and Random Sampling from Compact
Geographical Strata by k-Means.''} \emph{Computers \& Geosciences} 36
(10): 1261--67. \url{https://doi.org/10.1016/j.cageo.2010.04.005}.

\leavevmode\vadjust pre{\hypertarget{ref-wenger2012}{}}%
Wenger, Seth J., and Julian D. Olden. 2012. {``Assessing Transferability
of Ecological Models: An Underappreciated Aspect of Statistical
Validation.''} \emph{Methods in Ecology and Evolution} 3 (2): 260--67.
\url{https://doi.org/10.1111/j.2041-210x.2011.00170.x}.

\leavevmode\vadjust pre{\hypertarget{ref-wickham2019}{}}%
Wickham, Hadley, Mara Averick, Jennifer Bryan, Winston Chang, Lucy
McGowan, Romain François, Garrett Grolemund, et al. 2019. {``Welcome to
the Tidyverse.''} \emph{Journal of Open Source Software} 4 (43): 1686.
\url{https://doi.org/10.21105/joss.01686}.

\leavevmode\vadjust pre{\hypertarget{ref-ranger}{}}%
Wright, Marvin N., and Andreas Ziegler. 2017. {``{ranger}: A Fast
Implementation of Random Forests for High Dimensional Data in {C++} and
{R}.''} \emph{Journal of Statistical Software} 77 (1): 1--17.
\url{https://doi.org/10.18637/jss.v077.i01}.

\leavevmode\vadjust pre{\hypertarget{ref-yates2018}{}}%
Yates, Katherine L., Phil J. Bouchet, M. Julian Caley, Kerrie Mengersen,
Christophe F. Randin, Stephen Parnell, Alan H. Fielding, et al. 2018.
{``Outstanding Challenges in the Transferability of Ecological
Models.''} \emph{Trends in Ecology \& Evolution} 33 (10): 790--802.
\url{https://doi.org/10.1016/j.tree.2018.08.001}.

\end{CSLReferences}

\end{document}